\documentclass[twocolumn]{aastex62}

\newcommand{\highz}{high-$z$}
\newcommand{\aips}{\textsc{aips}}
\graphicspath{{./}{figures/}}

\received{January 1, 2018}
\revised{January 7, 2018}
\accepted{\today}
\submitjournal{ApJ}

\shorttitle{Radio emission from Blueberry galaxies}
\shortauthors{Sebastian \& Bait}

\begin{document}

\title{Radio continuum emission from local analogs of \highz{} faint LAEs: Blueberry galaxies}

\correspondingauthor{Biny Sebastian}
\email{biny@ncra.tifr.res.in}

\author[0000-0002-0786-7307]{Biny Sebastian}

\affil{National Centre for Radio Astrophysics, Tata Institute of Fundamental Research, Post Bag 3, Ganeshkhind, Pune 411007, India}

\author[0000-0003-2722-8841]{Omkar Bait}
\affiliation{National Centre for Radio Astrophysics, Tata Institute of Fundamental Research, Post Bag 3, Ganeshkhind, Pune 411007, India}



\begin{abstract}
We present a radio continuum study of a population of extremely young and starburst galaxies, termed as blueberries at $\sim$ 1 GHz using the upgraded Giant Metrewave Radio Telescope (uGMRT).
We find that their radio-based star formation rate (SFR) is suppressed by a factor of $\sim$ 3.4 compared to the SFR based on optical emission lines. This might be due to (i) the young ages of these galaxies as a result of which a stable equilibrium via feedback from supernovae has not yet been established (ii) escape of cosmic ray electrons via diffusion or galactic scale outflows. The estimated non-thermal fraction in these galaxies has a median value of $\sim$ 0.49, which is relatively lower than that in normal star-forming galaxies at such low frequencies. Their inferred equipartition magnetic field has a median value of 27 $\mu $G, which is higher than those in more evolved systems like spiral galaxies. Such high magnetic fields suggest that small-scale dynamo rather than large-scale dynamo mechanisms might be playing a major role in amplifying magnetic fields in these galaxies. 
\end{abstract}



\section{Introduction}
In the hierarchical galaxy formation model, dwarf galaxies are the first galaxies to have formed stars in the Universe. These galaxies are then thought to have undergone various events of mergers and accretion which transformed them into the massive galaxies we observe today.
Of these high redshift galaxy populations, Lyman-$\alpha$ emitters (LAEs) are of particular interest. Most of the \highz{} LAEs are star-forming galaxies with low mass, low metallicity, low dust extinction, young stellar population ($\leq$ 100 Myr) and compact sizes \cite[e.g.,][]{gawiser2007, malhotra2012}. Studies have also shown that faint dwarf starburst galaxies, a subset of which could be LAEs, can leak a significant amount of Lyman continuum (LyC) photons which could be major contributors for reionization \citep[e.g.,][]{Dressler15, Drake17}. Thus it is important to do a statistical study on the complex processes of star formation (SF), gas accretion and feedback in these \highz{} LAEs. But owing to their high redshifts it can be extremely difficult to study them in detail.

Due to these limitations, it is very useful to search for local analogs of \highz{} LAEs e.g., the bright green and compact (typically $\leq 5$ kpc)  galaxies termed as ``green pea" galaxies \citep{cardamone2009} discovered by the volunteers from the Galaxy Zoo project. These galaxies have extremely high SFRs, low stellar masses, low metallicities and have a bright green color due to the extremely strong [OIII] emission line. Green pea galaxies are also known to leak a significant amount of LyC photons which can ionize the surrounding intergalactic medium \citep[e.g.,][]{Izotov16}. \citet{yang2017} has searched for low redshift counterparts of the green-pea galaxies in the Sloan Digital Sky Survey (SDSS) $ugriz$ broadband images. They have found a sample of 40, spectroscopically confirmed, starburst dwarf galaxies with even smaller sizes ($\leq 1$ kpc), lower stellar masses and low metallicities, termed as ``Blueberry" galaxies. They are supposed to be the local analogs of \highz{} faint LAEs \citep{yang2017}.

A complementary study of this peculiar population of galaxies at radio frequencies can provide an independent estimate of their total (dust-obscured and unobscured) SFRs \citep[e.g.,][]{condon1992, Magnelli15, Pannella15, Bera18}. 
Radio emission from high redshift ($z \sim 3-4$) Lyman-break galaxies (LBGs) has been reported only by stacking analysis, \citep[e.g.,][]{carilli2008, Ho2010, To2014}. This stacked radio luminosity was also found to be systematically lower than those expected from the ultra-violet (UV) luminosity \citep{carilli2008}. However, for such galaxies, it is hard to separate the effect of an increased inverse Compton (IC) cooling of the relativistic electrons due to scattering from cosmic microwave background (CMB) photons versus an intrinsic deficit. The local analogs of LBGs and LAEs will have a negligible amount of IC-CMB losses and are also easier to detect. However, in a previous study on green peas, a direct detection in radio was found only for two of them using the Giant Metrewave Radio Telescope (GMRT) \citep{chakraborti2012}. 

In this Letter, we push these radio studies to the even younger and lower mass population of blueberry galaxies. We report their first ever radio detections, which has been possible due to the wideband backend of the upgraded GMRT \citep[uGMRT,][]{gupta17} allowing us to reach an rms of $\sim$ 15 $\mu$Jy beam$^{-1}$ with reasonable integration time at 1.25 GHz (Section~\ref{sec:obs}). Here we compare their radio-based SFRs with the optical SFR and estimate their non-thermal fraction and equipartition magnetic field strengths.

Throughout this paper, we use the standard concordance
cosmology from WMAP9 \citep{Hinshaw13} with $\Omega_M = 0.286$, $\Omega_\Lambda = 0.714$ and $h_{100} =
0.69$.

\section{GMRT Observations } \label{sec:obs}

\begin{table*}
 \centering
\begin{center}
\caption {\bf Sample and observation details }
\label{tab1}
\begin{tabular}{|p{0.5in}|p{1in}|p{1in}|p{0.5in}|p{0.5in}|p{0.7in}|p{1.0in}|p{0.6in}|p{0.4in}|}
\hline

ObjID$^a$ & R.A. & Decl. & Redshift$^b$ & Time on& Flux density & Beam Size & Beam P.A. \\
 & (deg)& (deg)& $z$ & source (min)& ($\mu$Jy) &  ($\arcsec \times \arcsec$) &($\degr$ )\\
\hline

27 & 01h46m53.307s & $+$03d19m22.360s & 0.047 &     239 & 105$\pm$15 &  6.87$\arcsec \times$ 3.24$\arcsec$&$ $71.96$\degr$  \\  
34 & 03h57m13.516s & $+$18d08m45.582s & 0.037 &     239 & 70 $\pm$14 &  3.04$\arcsec \times$ 2.48$\arcsec$&$ $88.50$\degr$ \\
3     & 08h25m40.449s & $+$18d46m17.209s & 0.038 &  214 & 72 $\pm$11 &  3.30$\arcsec \times$ 2.51$\arcsec$&$ $77.96$\degr$  \\   
5     & 10h32m56.727s & $+$49d19m47.226s & 0.044 &   183 & $<$39$^c$ &  4.23$\arcsec \times$ 2.40$\arcsec$&$ $75.81$\degr$  \\  
66 & 11h13m12.241s & $+$03d01m12.831s & 0.023 &     199 & 56 $\pm$15 &  3.39$\arcsec \times$ 2.42$\arcsec$&$-$10.31$\degr$  \\  
67 & 11h23m48.949s & $+$20d50m31.297s & 0.033 &    225 & 373 $\pm$17 &  3.88$\arcsec \times$ 2.31$\arcsec$&$ $87.74$\degr$  \\  
6     & 13h23m47.462s & $-$01d32m52.008s & 0.023 &   211 & 91 $\pm$18 & 2.59$\arcsec \times$ 2.43$\arcsec$&$ $49.60$\degr$  \\   
10 & 15h09m34.173s & $+$37d31m46.117s & 0.033 &    220 & 345 $\pm$16 &  3.14$\arcsec \times$ 2.55$\arcsec$&$-$69.71$\degr$  \\
12 & 15h56m24.474s & $+$48d06m45.792s & 0.050 &    241 & 107 $\pm$12 &  3.20$\arcsec \times$ 2.36$\arcsec$&$-$41.99$\degr$  \\  
13 & 16h08m10.363s & $+$35d28m09.346s & 0.033 &     261 & 58 $\pm$17 &  7.32$\arcsec \times$ 3.31$\arcsec$&$ $55.38$\degr$  \\  
\hline
\end{tabular}
\\
{$^a$ - The ObjIDs correspond to those from \cite{yang2017} } \\
{$^b$ - The redshifts are taken from \cite{yang2017} } \\
{$^c$ - 3$\sigma$ upper limit on the flux density}

\end{center}
\end{table*}
\subsection{Sample}
In this pilot study, we selected a subset of the 40 confirmed blueberry galaxies from \cite{yang2017} as follows. 
The SFR from \cite{yang2017} were used to calculate the expected non-thermal flux density at 1.4 GHz for a normal star-forming galaxy following \cite{yun2002}. 
Since this calibration can vary for our population of young starburst galaxies, we selected 10 of the brightest sources to ensure direct detections.

\begin{figure*}[htb!]

\centerline{
\includegraphics[width=17cm]{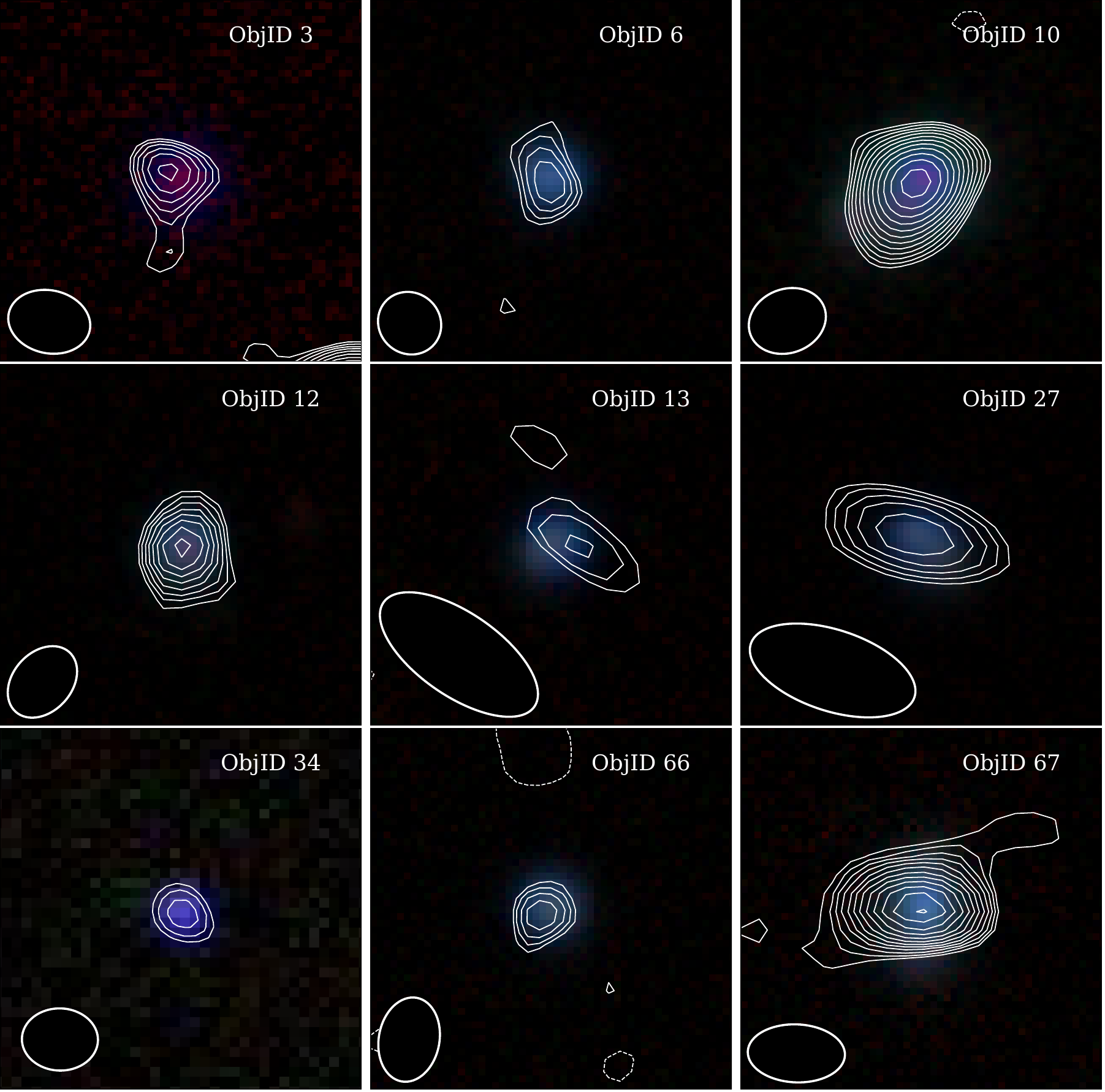}}
        
\caption{Contours of the 1.25 GHz continuum image overlaid on the optical $grz$-band color composite image (14.4 $\arcsec \times$ 14.4 $\arcsec$ in size) from DECaLS or MzLS (for sources without DECaLS data) or SDSS (for sources without DECaLS or MzLS data). The contour levels represent 3 $\sigma \times$($-$1.0, 1.0, 1.19, 1.41, 1.68, 2.0, 2.38, 2.83, 3.36, 4.0, 4.76, 5.66, 6.73, 8.0, 9.51, 11.31) $\mu$Jy beam $^{-1}$ where  $\sigma$ = 11.0, 13.0, 18.0, 16.0, 12.0, 14.3, 15.0, 14.0, 10.0, 20.0 for each of the sources sorted in the ascending order of their ObjIDs.  The clean beam is shown in the bottom left corner of each cutout.} 
\label{radio_overlay}
\end{figure*}

These galaxies were observed using uGMRT (project ID: 34$\_$123) at Band-5 (1000-1450 MHz).
The details of the observations are listed in Table~\ref{tab1}.
\subsection{Data Analysis}
Basic editing and calibration of the data were carried out using standard procedures in \aips{}. Data corrupted by RFI were identified and removed using automatic flagging algorithms in \aips{}. We carried out flux, phase and bandpass calibration using standard calibrators. 
 The flux calibrators used were 3C\,147, 3C\,286 and 3C\,48. The time resolution used for this observation was 10.1s and no further time averaging was carried out. Few channels were then averaged using task `SPLAT' in \aips{} which resulted in a final channel width of $\sim$ 3.5 MHz. This channel width would lead to 3\% reduction in the flux 5$\arcmin$ away from the centre of the field due to bandwidth smearing, whereas at the location of our target, there will only be negligible errors.
The calibrated target source was then extracted out of the main data set. Initial rounds of imaging and self-calibration were carried out in \aips{} using only a central bandwidth of 100 MHz to make sure bandwidth effects do not adversely affect the image quality.
Four rounds of the phase-only self-calibration were carried out in \aips{} to remove antenna based ionospheric errors. 
The target source was then exported to CASA and imaged using MS-MFS algorithm \citep{rau2011}. The flux densities of the target sources were then estimated using `JMFIT' task in \aips{}. The flux density values of all the sources in our sample are summarized in Table~\ref{tab1}. The errors on flux density quoted in Table~\ref{tab1} represent the rms from the images. However, for all the rest of the calculations we add 10 $\%$ calibration error on the flux density in quadrature. We found detections for 9 of the 10 sources in our sample. Figure \ref{radio_overlay} shows the contours of radio images of these nine sources overlaid on the $grz$-band color composite image from the Dark Energy Camera Legacy Survey (DECaLS) or the Mayall z-band Legacy Survey (MzLS) \citep[]{Dey19}. We used the $gri$-bands from SDSS \citep{abazajian2009} in the case of the source, ObjID-34 due to the unavailability of DECaLS or MzLS images.
\section{Results} \label{sec:results}

\begin{figure}

\centerline{
\includegraphics[width=9cm]{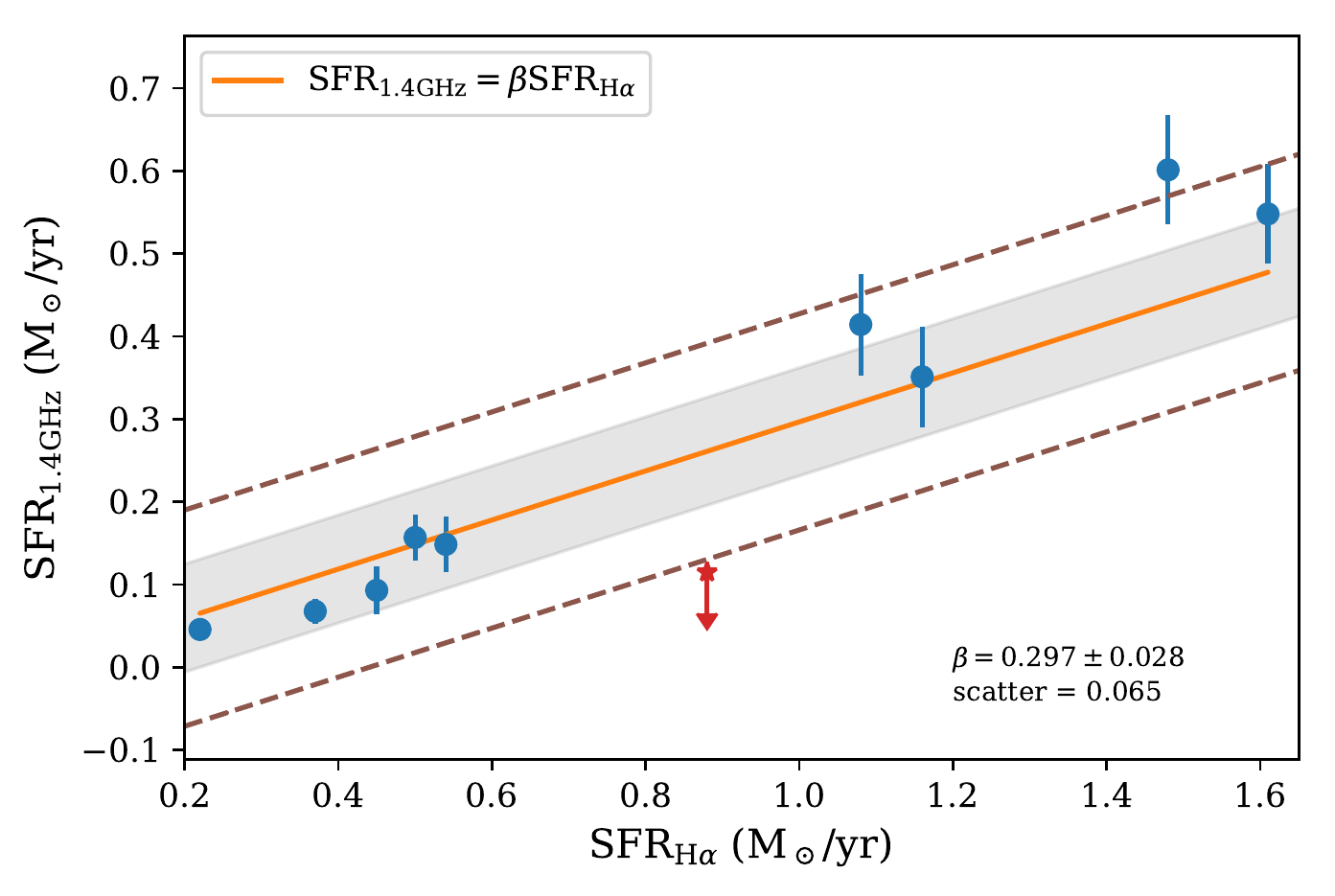}}

\caption{Comparison of SFR from H$\alpha$ and 1.4 GHz radio luminosity shown in blue points. The red star corresponds to the three sigma upper limit on SFR for one of the non-detection in our sample. The least square fit between the two SFRs (yellow solid line) has a slope of 0.297$\pm$0.028. The scatter in the residual, obtained after subtracting the best-fit line,  is 0.065 (gray shaded region). The dashed line shows the $2\sigma$ scatter of the residual. }
\label{fig:1}
\end{figure}

Normal star-forming galaxies show a tight correlation between the SFR and non-thermal radio emission and typically have a negligible amount of thermal emission at frequencies of $\sim$ 1 GHz and below \citep[$ \sim 10\% \pm 9 \%$, e.g.,][]{tabatabaei2017}. 
Here we investigate the SFR derived from the radio continuum emission for our sample of the blueberry galaxies and compare it with the SFR derived using optical emission lines. 

The radio-based SFRs for our sample is derived using the 1.4 GHz luminosity calibration from \citet{murphy2011} Eq. 17. For our sample of blueberries, the median value of the SFR from this calibrator is 0.15 M$_{\odot}$yr$^{-1}$. In Figure~\ref{fig:1} we plot the radio SFR against the SFR derived using the H$\alpha$ line from \citet{yang2017}. Although there is a correlation between the two SFRs, the star formation rates estimated from the radio emission are significantly lower than the ones deduced from H$\alpha$. 
Both the SFR calibrators use Kroupa IMF with the same stellar mass limits (0.1 M$_\odot$- 100 M$_\odot$). Hence we do not expect any difference in the two SFRs due to a difference in the IMF and/or it's adopted stellar mass range. Note that here and in all other further analysis we do not use ObjID 5 (red star) since it was not detected in our observation and hence does not have a reliable flux measurement. We do a least square fit between the SFR$_{1.4\mathrm{GHz}}$ and SFR$_{\mathrm{H}\alpha}$ which has a slope of 0.297$\pm$0.028. This factor can be used to correct the radio-based SFRs of blueberries. The scatter in the residual after removing the best-fit line is 0.065. Note that the 3$\sigma$ upper limit of the undetected source, ObjID 5 also lies within the 2$\sigma$ scatter of this relation. Blueberries can be dust obscured \citep{rong18}, which could mean that the SFR$_{\mathrm{H}\alpha}$ needs some dust correction. However, it will only further enhance the radio-suppression observed here.

The stacking detection of Green pea galaxies with VLA FIRST data showed suppression of radio continuum emission by a factor of $ \sim 0.53$ \citep{chakraborti2012}, however, the individual detections were closer to the expected values. \cite{carilli2008} also find a depression in the stacked radio continuum emission for a sample of \highz{} LBGs. 
They proposed that such a reduction can be a result of increased IC-CMB losses at \highz{} or a combination of IC-CMB losses and intrinsic reduction in radio continuum. Our sample of blueberry galaxies are at very low redshifts and the effects of IC-CMB is negligible which implies that the suppression is intrinsic. And it is possible that even the \highz{} LBGs show an intrinsic suppression of radio continuum.



One reason for the diminished radio emission might be the young age itself.
\cite{greis2017} point out that the standard SFR calibrators depend upon the stellar population age and star formation history. 
H$\alpha$ emission traces the recent star formation ($\sim 10$ Myrs), whereas the synchrotron or non-thermal radio emission traces star formation which has been continuing for about 100 Myrs. 
Another reason for the diminished radio emission could be that the relativistic electrons produced in the galaxy escape from these galaxies easily compared to other massive systems due to their shallow gravitational potential \citep{greis2017}. 

In both the above cases, only the non-thermal radio emission will show a decrement in the flux densities. Hence, the thermal radio emission should be consistent with that deduced using emission lines. \cite{tabatabaei2017} provides a calibration for converting the thermal radio emission to SFR using the KINGFISH galaxy sample \citep{kennicutt2011}. 
Under this assumption, we can estimate the non-thermal fraction as follows. Let $\delta$ be thermal fraction such that the thermal luminosity at L-band (L$_{th}$) is $\delta$ times the total observed radio luminosity at L-band (L$_{tot}$). And let $\beta$ be the thermal radio calibration factor used to estimate the SFR from \cite{tabatabaei2017}. We can then relate the SFR from H$\alpha$ emission and $\delta$ as: $SFR_{H \alpha} = SFR_{th} = \beta \times (L_{th}) = \ \beta \times (\delta \mathrm{L}_{tot})$. The non-thermal fraction, which is nothing but (1-$\delta$), can the be estimated using this above relation.

The median value of the non-thermal fraction for our sample of blueberry galaxies is $\sim 0.49$. Such a low non-thermal fraction was also previously observed in other types of dwarf galaxies e.g., in blue compact dwarfs \citep{Thuan2004, ramya2011}, in IC 10, a post-starbust dwarf irregular galaxy \citep{Hessen2011} and sample of faint star-forming dwarf galaxies \citep{sambit2012}. 
\cite{greis2017} showed that the fraction of thermal radio emission can be as high as $\sim 100\% $ for a sample of local Lyman break analogs (LBAs).
\begin{figure}

\centerline{
\includegraphics[width=9cm]{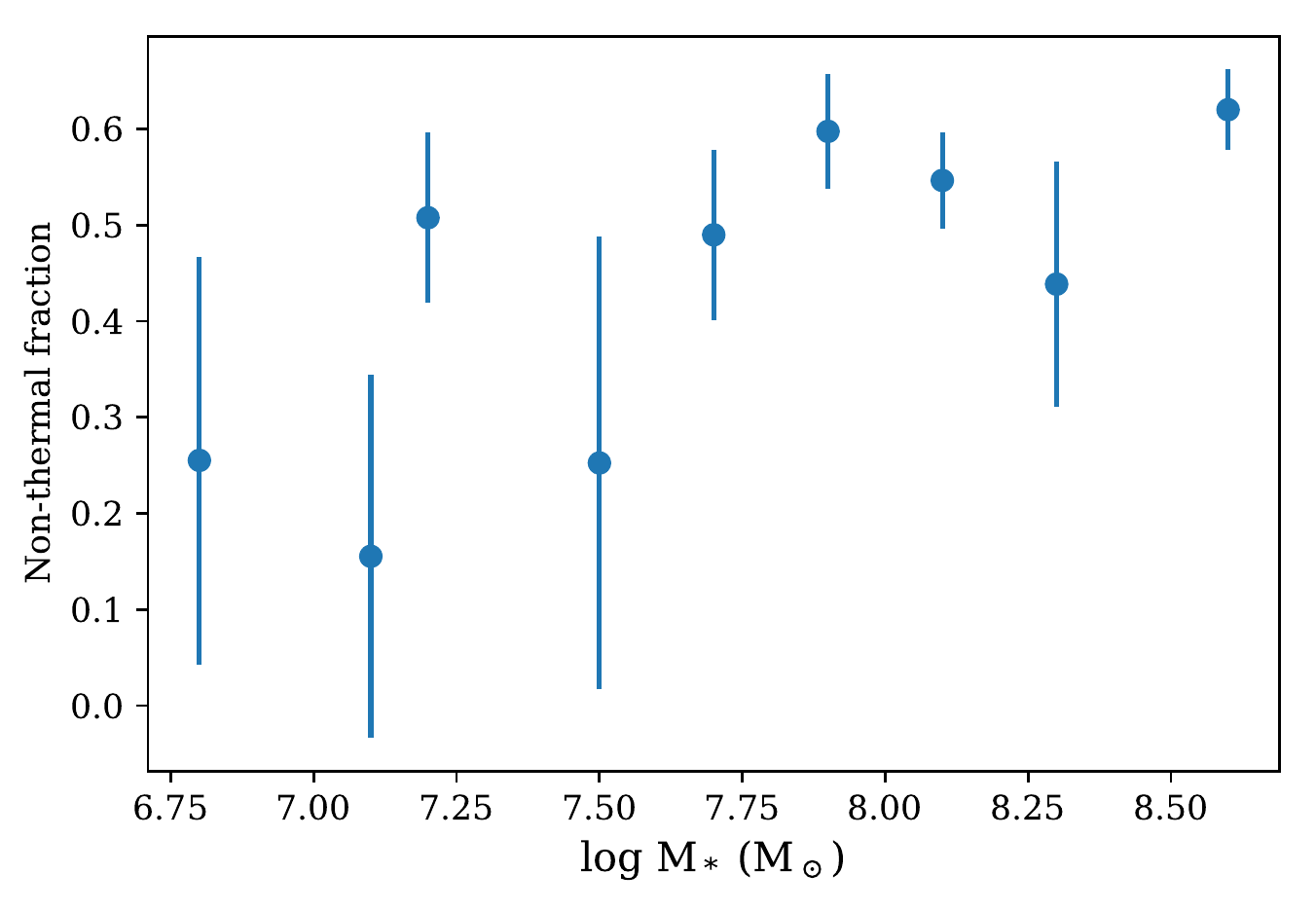}
}
\caption{Estimated non-thermal fraction as a function of stellar mass for our sample of blueberry galaxies.}
\label{fig:2}
\end{figure}

The difference in the non-thermal fraction in a sample of dwarf galaxies with that of normal galaxies suggests that there might be an evolution of the non-thermal fraction with the growth of galaxies, particularly in their stellar mass. We explore this in our sample by plotting the non-thermal fraction against the galaxy stellar mass in Figure~\ref{fig:2}. We see some correlation between the non-thermal fraction and stellar mass. The Spearman's rank correlation coefficient between the two variable is 0.68 (with a p-value of 0.04). However, given that the uncertainties in this fraction are large and our sample is small, we only claim a tentative correlation.

The origin of the non-thermal radio emission is attributed to the SNe shocks where the CREs are generated and the magnetic fields amplified. We estimate the magnetic fields from the non-thermal flux density values in these galaxies. We make use of the revised formula given by \cite{beck2005} for calculating the equipartition magnetic fields. We assume the ratio of the relativistic proton number density to that of the electrons to be equal to 100 and a spectral index value of $-$0.7. We assume a path-length of 0.3 kpc following \cite{chakraborti2012}. The median equipartition magnetic fields for our sample is estimated to be $\sim$ 27 $\mu$G.

\section{Discussion} \label{sec:discuss}

To identify the reason for the deficit and the evolution in the radio non-thermal emission we compare the characteristic timescales of various processes (age of the recent starburst, build-up of SN rate and CRE escape via diffusion or outflows).

\subsection{Young ages}
Assuming that most of the stellar mass was built up in the starburst phase, we can use the current SFR to put an upperlimit on the time elapsed since the onset of this burst. We find that this timescale ranges from 30 - 370 Myr with a median value of 70 Myr for our sample of galaxies. \cite{greis2017} show using a stellar population synthesis for a system which forms stars at a rate of 1 M$_{\odot}$ yr$^{-1}$ that the SFR traced by H$\alpha$ and [O II] becomes significant just after the star formation commences. Whereas the synchrotron emission takes about 10 Myr to become significant and matches with the optical-based SFRs only after 100 Myr after the onset of star formation. Thermal emission, on the other hand, traces the SFR on timescales similar to H$\alpha$ and [O II]. 
{}

 Even the evolution of the non-thermal fraction (see Figure~\ref{fig:2}) can be easily explained if the decrement is due to young age. The bolometric non-thermal emission, $L^{NT}$ from a galaxy can be roughly calculated as $L^{NT}=\nu_{CCSNe} \times E_{el}$ \citep{bressan2002},  where $\nu_{CCSN}$ is the rate of the occurrence of CCSNe and E$_{el}$ is the power emitted by an electron via synchrotron losses times the lifetime of the electron. There is an overall dependence of the non-thermal emission on the magnetic-field since E$_{el}$ also depends on the magnetic-field. \cite{greis2017} consider only the evolution in the rate of CCSNe which stabilizes after about 10 Myr but
the magnetic field could be evolving for a longer timescale.
\cite{schleicher2010} suggest that saturation of magnetic fields amplified via small-scale dynamo (SSD; which could be the preferred model for blueberries) happens only after $\sim$ 100 Myr.   Hence, the magnetic fields in our sample of blueberry galaxies might still be in the evolutionary phase as a result of which the non-thermal fraction shows an evolution.

 Thus in blueberry galaxies, which form one of the youngest class of star-forming galaxies (with median ages $\leq 70 Myr $), it is plausible that the star formation has not sustained long enough for synchrotron emission to become dominant.

\subsection{CRE escape via diffusion/outflows}
 The deficit in the non-thermal radio emission can also be explained by the escape of CREs via diffusion and/or winds/outflows. This happens only if their timescales are lower than the synchrotron-loss timescales \citep{carilli2008}. Diffusion timescale ($t_D$) can be estimated as $t_D=\frac{R^2}{D}$, where D, the diffusion coefficient is equal to $\frac{l_0c}{3}$ and R is the size of the galaxy. The typical values of the mean free path, $l_0$=0.3 kpc (see \cite{chakraborti2012} and \cite{calvez2010} for details). For an upper-limit of 1 kpc on the size we put an upper-limit of 0.033 Myr on $t_D$.  We estimate the synchrotron-loss timescales using the relation, $\tau_{syn}=33.4B_{10}^{-\frac{3}{2}}\nu_1^{-\frac{1}{2}}$ \citep{perez2007}, where $B_{10}$ is the magnetic field value in units of 10 $\mu$G, and $\nu_1$ is the spectral break frequency in GHz. For typical values of break frequencies ranging from 1 - 12 GHz \citep{klein2018}, the synchrotron-loss timescales turn out to be $\sim$2 - 7.5 Myr.
Since $t_D << \tau_{syn}$, CRE diffusion might be playing a major role in the suppression of radio continuum emission in these galaxies.


Another possibility is the escape of CREs due to winds or outflows.
Galactic superwinds were shown to be present in all the galaxies which had global SFR densities (SFRDs) greater than 0.1 M$_{\odot}$ yr$^{-1}$ kpc$^{-2}$ \citep{heckman2002}.
Assuming an upper limit of 1~kpc for the size of these blueberry galaxies, the SFRDs of all the galaxies are higher than 0.1 M$_{\odot}$ yr$^{-1}$ kpc$^{-2}$.
 The typical wind terminal velocities attained in starbursting galaxies hosting superwind are in the range, 2000 - 3000 km/s \citep{heckman1990}. Assuming a wind velocity of 2000 km/s, and a galaxy size of 1 kpc, we obtain escape timescales of 0.5 Myr, which is also lower than the synchrotron loss timescales.
Such an outflow would only be a natural consequence of the massive star formation that is happening in these young massive galaxies. 

In our study, we are not able to distinguish between the different scenarios which lead to the decrement in the radio continuum emission. 

\subsection{Amplification of magnetic-fields}

The magnetic field for our galaxies are much higher than those seen in more evolved systems like normal spiral galaxies \citep[e.g., about 9$\pm$2$\mu$G][]{niklas1995}. 

The magnetic field amplification is explained widely using the dynamo mechanism. 
The large-scale dynamo models can effectively convert the kinetic energy to magnetic energy in spiral galaxies where differential rotation of conducting fluid is present. 
However, blueberry galaxies are much younger and do not yet possess the kind of large scale differential rotation seen in spirals. 
Small scale dynamo (SSD) models, on the other hand, amplify magnetic field by converting turbulent energy due to supernova shocks \citep{ferriere1996} to the magnetic energy at much smaller scales. 
Moreover, the growth rates predicted by SSD models are larger compared to the large-scale dynamo models \citep{brandenburg2005}. 
Hence, the SSD mechanism is often invoked to explain the $\mu$G level magnetic fields observed in young galaxies \citep{kulsrud1997,rieder2015}.
Our study also favors SSD mechanism for the amplification of magnetic fields compared to the large-scale dynamo models. A study of the evolution of magnetic fields with the stellar mass in such systems can provide useful insights into the underlying mechanisms that amplify the magnetic fields. \\ 

In the future, we aim to do a detailed radio spectral modeling \citep{klein2018} to make accurate estimates of the thermal/non-thermal fraction and as a result, the magnetic field. In the case where CREs escape the galaxy and causes the decrement in the radio emission, it would show a break in the radio synchrotron spectrum \citep{lisenfeld2004,klein2018}. We would also increase the sample size to include both the lower mass and higher mass galaxies (green pea galaxies) to study the evolution of the non-thermal properties in these systems. We also aim to do a detailed comparison of SFRs from different indicators (e.g., UV+IR, H$\alpha$, radio flux density, and multi-wavelength spectral energy distribution fitting).

\acknowledgements
We thank the anonymous referee for her/his insightful comments that helped in improving the article significantly. We also thank S. Kurapati, A. Bera and S. Manna for several of their suggestions and comments. We thank the staff of the GMRT who have made these observations possible. The GMRT is run by the National Center for Radio Astrophysics of the Tata Institute of Fundamental Research.


\begin{thebibliography}{}
\expandafter\ifx\csname natexlab\endcsname\relax\def\natexlab#1{#1}\fi
\providecommand{\url}[1]{\href{#1}{#1}}

\bibitem[{{Abazajian} {et~al.}(2009){Abazajian}, {Adelman-McCarthy},
  {Ag{\"u}eros}, {Allam}, {Allende Prieto}, {An}, {Anderson}, {Anderson},
  {Annis}, {Bahcall}, \& et~al.}]{abazajian2009}
{Abazajian}, K.~N., {Adelman-McCarthy}, J.~K., {Ag{\"u}eros}, M.~A., {et~al.}
  2009, \apjs, 182, 543

\bibitem[{{Beck} \& {Krause}(2005)}]{beck2005}
{Beck}, R., \& {Krause}, M. 2005, Astronomische Nachrichten, 326, 414

\bibitem[{{Bera} {et~al.}(2018){Bera}, {Kanekar}, {Weiner}, {Sethi}, \&
  {Dwarakanath}}]{Bera18}
{Bera}, A., {Kanekar}, N., {Weiner}, B.~J., {Sethi}, S., \& {Dwarakanath},
  K.~S. 2018, \apj, 865, 39

\bibitem[{{Brandenburg} \& {Subramanian}(2005)}]{brandenburg2005}
{Brandenburg}, A., \& {Subramanian}, K. 2005, \physrep, 417, 1

\bibitem[{{Bressan} {et~al.}(2002){Bressan}, {Silva}, \&
  {Granato}}]{bressan2002}
{Bressan}, A., {Silva}, L., \& {Granato}, G.~L. 2002, \aap, 392, 377

\bibitem[{{Calvez} {et~al.}(2010){Calvez}, {Kusenko}, \&
  {Nagataki}}]{calvez2010}
{Calvez}, A., {Kusenko}, A., \& {Nagataki}, S. 2010, \prl, 105, 091101

\bibitem[{{Cardamone} {et~al.}(2009){Cardamone}, {Schawinski}, {Sarzi},
  {Bamford}, {Bennert}, {Urry}, {Lintott}, {Keel}, {Parejko}, {Nichol},
  {Thomas}, {Andreescu}, {Murray}, {Raddick}, {Slosar}, {Szalay}, \&
  {Vandenberg}}]{cardamone2009}
{Cardamone}, C., {Schawinski}, K., {Sarzi}, M., {et~al.} 2009, \mnras, 399,
  1191

\bibitem[{{Carilli} {et~al.}(2008){Carilli}, {Lee}, {Capak}, {Schinnerer},
  {Lee}, {McCraken}, {Yun}, {Scoville}, {Smol{\v c}i{\'c}}, {Giavalisco},
  {Datta}, {Taniguchi}, \& {Urry}}]{carilli2008}
{Carilli}, C.~L., {Lee}, N., {Capak}, P., {et~al.} 2008, \apj, 689, 883

\bibitem[{{Chakraborti} {et~al.}(2012){Chakraborti}, {Yadav}, {Cardamone}, \&
  {Ray}}]{chakraborti2012}
{Chakraborti}, S., {Yadav}, N., {Cardamone}, C., \& {Ray}, A. 2012, \apjl, 746,
  L6

\bibitem[{{Condon}(1992)}]{condon1992}
{Condon}, J.~J. 1992, \araa, 30, 575

\bibitem[{{Dey} {et~al.}(2019){Dey}, {Schlegel}, {Lang}, {Blum}, {Burleigh},
  {Fan}, {Findlay}, {Finkbeiner}, {Herrera}, {Juneau}, {Landriau}, {Levi},
  {McGreer}, {Meisner}, {Myers}, {Moustakas}, {Nugent}, {Patej}, {Schlafly},
  {Walker}, {Valdes}, {Weaver}, {Y{\`e}che}, {Zou}, {Zhou}, {Abareshi},
  {Abbott}, {Abolfathi}, {Aguilera}, {Alam}, {Allen}, {Alvarez}, {Annis},
  {Ansarinejad}, {Aubert}, {Beechert}, {Bell}, {BenZvi}, {Beutler}, {Bielby},
  {Bolton}, {Brice{\~n}o}, {Buckley-Geer}, {Butler}, {Calamida}, {Carlberg},
  {Carter}, {Casas}, {Castander}, {Choi}, {Comparat}, {Cukanovaite}, {Delubac},
  {DeVries}, {Dey}, {Dhungana}, {Dickinson}, {Ding}, {Donaldson}, {Duan},
  {Duckworth}, {Eftekharzadeh}, {Eisenstein}, {Etourneau}, {Fagrelius},
  {Farihi}, {Fitzpatrick}, {Font-Ribera}, {Fulmer}, {G{\"a}nsicke},
  {Gaztanaga}, {George}, {Gerdes}, {Gontcho}, {Gorgoni}, {Green}, {Guy},
  {Harmer}, {Hernandez}, {Honscheid}, {Huang}, {James}, {Jannuzi}, {Jiang},
  {Joyce}, {Karcher}, {Karkar}, {Kehoe}, {Kneib}, {Kueter-Young}, {Lan},
  {Lauer}, {Le Guillou}, {Le Van Suu}, {Lee}, {Lesser}, {Perreault Levasseur},
  {Li}, {Mann}, {Marshall}, {Mart{\'{\i}}nez-V{\'a}zquez}, {Martini}, {du Mas
  des Bourboux}, {McManus}, {Meier}, {M{\'e}nard}, {Metcalfe},
  {Mu{\~n}oz-Guti{\'e}rrez}, {Najita}, {Napier}, {Narayan}, {Newman}, {Nie},
  {Nord}, {Norman}, {Olsen}, {Paat}, {Palanque-Delabrouille}, {Peng},
  {Poppett}, {Poremba}, {Prakash}, {Rabinowitz}, {Raichoor}, {Rezaie},
  {Robertson}, {Roe}, {Ross}, {Ross}, {Rudnick}, {Safonova}, {Saha},
  {S{\'a}nchez}, {Savary}, {Schweiker}, {Scott}, {Seo}, {Shan}, {Silva},
  {Slepian}, {Soto}, {Sprayberry}, {Staten}, {Stillman}, {Stupak}, {Summers},
  {Sien Tie}, {Tirado}, {Vargas-Maga{\~n}a}, {Vivas}, {Wechsler}, {Williams},
  {Yang}, {Yang}, {Yapici}, {Zaritsky}, {Zenteno}, {Zhang}, {Zhang}, {Zhou}, \&
  {Zhou}}]{Dey19}
{Dey}, A., {Schlegel}, D.~J., {Lang}, D., {et~al.} 2019, \aj, 157, 168

\bibitem[{{Drake} {et~al.}(2017){Drake}, {Garel}, {Wisotzki}, {Leclercq},
  {Hashimoto}, {Richard}, {Bacon}, {Blaizot}, {Caruana}, {Conseil}, {Contini},
  {Guiderdoni}, {Herenz}, {Inami}, {Lewis}, {Mahler}, {Marino}, {Pello},
  {Schaye}, {Verhamme}, {Ventou}, \& {Weilbacher}}]{Drake17}
{Drake}, A.~B., {Garel}, T., {Wisotzki}, L., {et~al.} 2017, \aap, 608, A6

\bibitem[{{Dressler} {et~al.}(2015){Dressler}, {Henry}, {Martin}, {Sawicki},
  {McCarthy}, \& {Villaneuva}}]{Dressler15}
{Dressler}, A., {Henry}, A., {Martin}, C.~L., {et~al.} 2015, \apj, 806, 19

\bibitem[{{Ferri{\`e}re} \& {Blanc}(1996)}]{ferriere1996}
{Ferri{\`e}re}, K.~M., \& {Blanc}, M. 1996, \jgr, 101, 19871

\bibitem[{{Gawiser} {et~al.}(2007){Gawiser}, {Francke}, {Lai}, {Schawinski},
  {Gronwall}, {Ciardullo}, {Quadri}, {Orsi}, {Barrientos}, {Blanc}, {Fazio},
  {Feldmeier}, {Huang}, {Infante}, {Lira}, {Padilla}, {Taylor}, {Treister},
  {Urry}, {van Dokkum}, \& {Virani}}]{gawiser2007}
{Gawiser}, E., {Francke}, H., {Lai}, K., {et~al.} 2007, \apj, 671, 278

\bibitem[{{Greis} {et~al.}(2017){Greis}, {Stanway}, {Levan}, {Davies}, \&
  {Eldridge}}]{greis2017}
{Greis}, S.~M.~L., {Stanway}, E.~R., {Levan}, A.~J., {Davies}, L.~J.~M., \&
  {Eldridge}, J.~J. 2017, \mnras, 470, 489

\bibitem[{Gupta {et~al.}(2017)Gupta, Ajithkumar, Kale, Nayak, Sabhapathy,
  Sureshkumar, Swami, Chengalur, Ghosh, Ishwara-Chandra, Joshi, Kanekar, Lal,
  \& Roy}]{gupta17}
Gupta, Y., Ajithkumar, B., Kale, H., {et~al.} 2017, Current Science, 113, 707

\bibitem[{{Heckman}(2002)}]{heckman2002}
{Heckman}, T.~M. 2002, in Astronomical Society of the Pacific Conference
  Series, Vol. 254, Extragalactic Gas at Low Redshift, ed. J.~S. {Mulchaey} \&
  J.~T. {Stocke}, 292

\bibitem[{{Heckman} {et~al.}(1990){Heckman}, {Armus}, \& {Miley}}]{heckman1990}
{Heckman}, T.~M., {Armus}, L., \& {Miley}, G.~K. 1990, \apjs, 74, 833

\bibitem[{{Heesen} {et~al.}(2011){Heesen}, {Rau}, {Rupen}, {Brinks}, \&
  {Hunter}}]{Hessen2011}
{Heesen}, V., {Rau}, U., {Rupen}, M.~P., {Brinks}, E., \& {Hunter}, D.~A. 2011,
  \apjl, 739, L23

\bibitem[{{Hinshaw} {et~al.}(2013){Hinshaw}, {Larson}, {Komatsu}, {Spergel},
  {Bennett}, {Dunkley}, {Nolta}, {Halpern}, {Hill}, {Odegard}, {Page}, {Smith},
  {Weiland}, {Gold}, {Jarosik}, {Kogut}, {Limon}, {Meyer}, {Tucker}, {Wollack},
  \& {Wright}}]{Hinshaw13}
{Hinshaw}, G., {Larson}, D., {Komatsu}, E., {et~al.} 2013, The Astrophysical
  Journal Supplement Series, 208, 19

\bibitem[{{Ho} {et~al.}(2010){Ho}, {Wang}, {Morrison}, \& {Miller}}]{Ho2010}
{Ho}, I.-T., {Wang}, W.-H., {Morrison}, G.~E., \& {Miller}, N.~A. 2010, \apj,
  722, 1051

\bibitem[{{Izotov} {et~al.}(2016){Izotov}, {Orlitov{\'a}}, {Schaerer}, {Thuan},
  {Verhamme}, {Guseva}, \& {Worseck}}]{Izotov16}
{Izotov}, Y.~I., {Orlitov{\'a}}, I., {Schaerer}, D., {et~al.} 2016, \nat, 529,
  178

\bibitem[{{Kennicutt} {et~al.}(2011){Kennicutt}, {Calzetti}, {Aniano},
  {Appleton}, {Armus}, {Beir{\~a}o}, {Bolatto}, {Brandl}, {Crocker}, {Croxall},
  {Dale}, {Donovan Meyer}, {Draine}, {Engelbracht}, {Galametz}, {Gordon},
  {Groves}, {Hao}, {Helou}, {Hinz}, {Hunt}, {Johnson}, {Koda}, {Krause},
  {Leroy}, {Li}, {Meidt}, {Montiel}, {Murphy}, {Rahman}, {Rix}, {Roussel},
  {Sandstrom}, {Sauvage}, {Schinnerer}, {Skibba}, {Smith}, {Srinivasan},
  {Vigroux}, {Walter}, {Wilson}, {Wolfire}, \& {Zibetti}}]{kennicutt2011}
{Kennicutt}, R.~C., {Calzetti}, D., {Aniano}, G., {et~al.} 2011, \pasp, 123,
  1347

\bibitem[{{Klein} {et~al.}(2018){Klein}, {Lisenfeld}, \& {Verley}}]{klein2018}
{Klein}, U., {Lisenfeld}, U., \& {Verley}, S. 2018, \aap, 611, A55

\bibitem[{{Kulsrud} {et~al.}(1997){Kulsrud}, {Cen}, {Ostriker}, \&
  {Ryu}}]{kulsrud1997}
{Kulsrud}, R.~M., {Cen}, R., {Ostriker}, J.~P., \& {Ryu}, D. 1997, \apj, 480,
  481

\bibitem[{{Lisenfeld} {et~al.}(2004){Lisenfeld}, {Wilding}, {Pooley}, \&
  {Alexander}}]{lisenfeld2004}
{Lisenfeld}, U., {Wilding}, T.~W., {Pooley}, G.~G., \& {Alexander}, P. 2004,
  \mnras, 349, 1335

\bibitem[{{Magnelli} {et~al.}(2015){Magnelli}, {Ivison}, {Lutz}, {Valtchanov},
  {Farrah}, {Berta}, {Bertoldi}, {Bock}, {Cooray}, {Ibar}, {Karim}, {Le
  Floc'h}, {Nordon}, {Oliver}, {Page}, {Popesso}, {Pozzi}, {Rigopoulou},
  {Riguccini}, {Rodighiero}, {Rosario}, {Roseboom}, {Wang}, \&
  {Wuyts}}]{Magnelli15}
{Magnelli}, B., {Ivison}, R.~J., {Lutz}, D., {et~al.} 2015, \aap, 573, A45

\bibitem[{{Malhotra} {et~al.}(2012){Malhotra}, {Rhoads}, {Finkelstein},
  {Hathi}, {Nilsson}, {McLinden}, \& {Pirzkal}}]{malhotra2012}
{Malhotra}, S., {Rhoads}, J.~E., {Finkelstein}, S.~L., {et~al.} 2012, \apjl,
  750, L36

\bibitem[{{Murphy} {et~al.}(2011){Murphy}, {Condon}, {Schinnerer}, {Kennicutt},
  {Calzetti}, {Armus}, {Helou}, {Turner}, {Aniano}, {Beir{\~a}o}, {Bolatto},
  {Brandl}, {Croxall}, {Dale}, {Donovan Meyer}, {Draine}, {Engelbracht},
  {Hunt}, {Hao}, {Koda}, {Roussel}, {Skibba}, \& {Smith}}]{murphy2011}
{Murphy}, E.~J., {Condon}, J.~J., {Schinnerer}, E., {et~al.} 2011, \apj, 737,
  67

\bibitem[{{Niklas}(1995)}]{niklas1995}
{Niklas}, S. 1995, PhD thesis, PhD Thesis, Univ.~Bonn, (1995)

\bibitem[{{Pannella} {et~al.}(2015){Pannella}, {Elbaz}, {Daddi}, {Dickinson},
  {Hwang}, {Schreiber}, {Strazzullo}, {Aussel}, {Bethermin}, {Buat},
  {Charmandaris}, {Cibinel}, {Juneau}, {Ivison}, {Le Borgne}, {Le Floc'h},
  {Leiton}, {Lin}, {Magdis}, {Morrison}, {Mullaney}, {Onodera}, {Renzini},
  {Salim}, {Sargent}, {Scott}, {Shu}, \& {Wang}}]{Pannella15}
{Pannella}, M., {Elbaz}, D., {Daddi}, E., {et~al.} 2015, \apj, 807, 141

\bibitem[{{P{\'e}rez-Torres} \& {Alberdi}(2007)}]{perez2007}
{P{\'e}rez-Torres}, M.~A., \& {Alberdi}, A. 2007, \mnras, 379, 275

\bibitem[{{Ramya} {et~al.}(2011){Ramya}, {Kantharia}, \& {Prabhu}}]{ramya2011}
{Ramya}, S., {Kantharia}, N.~G., \& {Prabhu}, T.~P. 2011, \apj, 728, 124

\bibitem[{{Rau} \& {Cornwell}(2011)}]{rau2011}
{Rau}, U., \& {Cornwell}, T.~J. 2011, \aap, 532, A71

\bibitem[{{Rieder} \& {Teyssier}(2015)}]{rieder2015}
{Rieder}, M., \& {Teyssier}, R. 2015, IAU General Assembly, 22, 2243950

\bibitem[{{Rong} {et~al.}(2018){Rong}, {Yang}, {Zhang}, {Puzia},
  {Chilingarian}, {Eigenthaler}, {Malhotra}, {Rhoads}, {Wang},
  {Ordens-briceno}, \& {Johnston}}]{rong18}
{Rong}, Y., {Yang}, H., {Zhang}, H.-x., {et~al.} 2018, arXiv e-prints,
  arXiv:1806.10149

\bibitem[{{Roychowdhury} \& {Chengalur}(2012)}]{sambit2012}
{Roychowdhury}, S., \& {Chengalur}, J.~N. 2012, \mnras, 423, L127

\bibitem[{{Schleicher} {et~al.}(2010){Schleicher}, {Banerjee}, {Sur},
  {Arshakian}, {Klessen}, {Beck}, \& {Spaans}}]{schleicher2010}
{Schleicher}, D.~R.~G., {Banerjee}, R., {Sur}, S., {et~al.} 2010, \aap, 522,
  A115

\bibitem[{{Tabatabaei} {et~al.}(2017){Tabatabaei}, {Schinnerer}, {Krause},
  {Dumas}, {Meidt}, {Damas-Segovia}, {Beck}, {Murphy}, {Mulcahy}, {Groves},
  {Bolatto}, {Dale}, {Galametz}, {Sandstrom}, {Boquien}, {Calzetti},
  {Kennicutt}, {Hunt}, {De Looze}, \& {Pellegrini}}]{tabatabaei2017}
{Tabatabaei}, F.~S., {Schinnerer}, E., {Krause}, M., {et~al.} 2017, \apj, 836,
  185

\bibitem[{{Thuan} {et~al.}(2004){Thuan}, {Hibbard}, \&
  {L{\'e}vrier}}]{Thuan2004}
{Thuan}, T.~X., {Hibbard}, J.~E., \& {L{\'e}vrier}, F. 2004, \aj, 128, 617

\bibitem[{{To} {et~al.}(2014){To}, {Wang}, \& {Owen}}]{To2014}
{To}, C.-H., {Wang}, W.-H., \& {Owen}, F.~N. 2014, \apj, 792, 139

\bibitem[{{Yang} {et~al.}(2017){Yang}, {Malhotra}, {Rhoads}, \&
  {Wang}}]{yang2017}
{Yang}, H., {Malhotra}, S., {Rhoads}, J.~E., \& {Wang}, J. 2017, \apj, 847, 38

\bibitem[{{Yun} \& {Carilli}(2002)}]{yun2002}
{Yun}, M.~S., \& {Carilli}, C.~L. 2002, \apj, 568, 88

\end{thebibliography}



\end{document}